\documentstyle[psfig]{mn2e}
\input{psfig.sty}
\newif\ifAMStwofonts

\title[Radio diagnostics of GRB properties]
      {Flaring up: radio diagnostics of the kinematic, hydrodynamic and
environmental properties of GRBs}
\author[ A.~M. Soderberg \& E. Ramirez-Ruiz]
       {Alicia M. Soderberg$^{1}$ and Enrico Ramirez-Ruiz\\
Institute of Astronomy, Madingley Road, Cambridge, CB3 0HA, UK.\\
$^{1}$Current address: Department of Astronomy, 105-24, California Institute of
      Technology, Pasadena, CA 91125, USA}
\date{}
\begin{document}
\maketitle
\label{firstpage}
\begin{abstract}
The specific incidence of radio flares appears to be significantly
larger than that of the prompt optical emission. This abundance,
coupled with the reverse shock interpretation suggests that radio
flares add a unique probe on the physics of GRB
shocks. Motivated thus, we estimate the strength of the reverse shock
expected for bursts in which multi-wavelength observations have
allowed the physical parameters of the forward shock to be
determined.  We use all 6 bursts
(980519, 990123, 990510, 991208, 991216, 000418) which are found to be
adiabatic and thus predicted to have a strong reverse shock.  
We aim to constrain the hydrodynamic evolution of the reverse
shock and the initial bulk Lorentz factor -- which we found to
be between $10^{2}$ and $10^{3}$ and well above the lower limits derived 
from the requirement that gamma-ray bursts be optically thin to high-energy
photons.
In half of the cases we improve the description of the early 
afterglow lightcurves 
by adding a contribution from the reverse
shock.
Modelling of this early emission provides the opportunity 
to investigate the 
immediate surroundings of the burst.  For 991216 and 991208, the expected
$1/r^2$ density structure for a stellar wind is not compatible with
the early afterglow lightcurves. Considering the
radial range relevant to these GRBs, we discuss
the conditions under which the inclusion of a wind termination shock may
resolve the absence of a $1/r^2$ density profile.
\end{abstract}
\section{Introduction}

The prompt and extremely bright optical flash in GRB 990123 (Akerlof
et al. 1999) was accompanied by a strong radio flare (Kulkarni et
al. 1999).  Peaking at $\sim$1 day, the flare was unlike usual radio
afterglows which rise to maximum on a timescale of weeks or even
months. The simplest interpretation of this radio flare is that it
arises from the reverse shock component of the external shock (Sari
$\&$ Piran 1999, hereinafter SP99). The reverse shock (hereinafter RS)
propagates into the adiabatically cooled particles of the coasting
ejecta, thereby shocking the shell material and producing a prompt
optical flash.  The RS emission then weakens rapidly and shifts
to lower energies, eventually crossing the observed radio band.
Indeed, it has been shown that the radio counterpart to GRB 990123 is
compatible with estimates derived from scaling down the prompt optical
emission to the epoch of the radio afterglow observations (Kulkarni et
al. 1999).

It turns out that there is a broad range of model parameters for which
a strong optical flash is expected to precede the main GRB afterglow (Soderberg
$\&$ Ramirez-Ruiz 2002; hereinafter SR02). Such flashes are, however,
difficult to detect in the current observing modes (Akerlof
et al. 2000, Paczy\'nski 2001). On the other hand, the specific
incidence of radio flares of 1:4 (Djorgovski et al. 2001) appears to
be significantly larger than that of the prompt optical emission
obtained by ROTSE (Kehoe et al. 2001) or LOTIS (Williams et
al. 2000). This abundance, together with the reverse shock
interpretation, suggests that the radio flare phenomenon has the
potential to shed new light on the physics of GRBs. Motivated by this
interpretation, we have estimated the strength of the radio flare
expected from bursts in which broadband observations have been able to
constrain the physical parameters of the forward shock (hereinafter
FS) emission (Panaitescu $\&$ Kumar 2001b; Panaitescu $\&$ Kumar 2002,
hereinafter PK02). We use all 6 bursts
(980519, 990123, 990510, 991208, 991216, 000418) which are found to be
adiabatic (i.e. $\nu_m < \nu_c$) for reasonable assumptions about
$\Gamma_0$ and thus predicted to have a strong RS.  
Radiative bursts (e.g. 970508) are not selected 
since the reverse shock component is significantly quenched
in this regime, and therefore the observations in these cases
are dominated entirely by emission from the forward shock. 
By evolving prompt optical flux estimates to the
epoch of radio afterglow observations, we are able to discern whether
a contribution from the RS could have been detected. We find that for
${1 \over 2}$ of the bursts for which the cooling frequency is larger
than the typical synchrotron frequency, the predicted RS emission (which is
generally found to lie in the mildly relativistic temperature regime)
combined with that of the FS produces an improved fit to the radio
afterglow data.  As we shall discuss, the strong dependence of the
peak time of this radio flare on the bulk Lorentz factor $\Gamma$
provides a way to measure this elusive parameter. Accepting the
conclusion that the radio flare arises from the RS, we then place
constraints on the velocity of source expansion and hence on the
density and profile of the medium in the immediate vicinity of six
bursts (980519, 990123, 990510, 991208, 991216, 000418). These
constraints provide environmental diagnostics for GRBs and are
compared with the main types of environments considered for
afterglows.  We assume $H_0 =65\,\, {\rm km} \, {\rm s}^{-1} \, {\rm
Mpc}^{-1}$, $\Omega_{\rm M}=0.3$, and $\Omega_{\Lambda}=0.7$.

\section{The reverse shock: optical flash to radio flare}

Observations of the optical flash associated with GRB 990123 confirmed
earlier predictions of prompt emission from a RS (M\'esz\'aros $\&$
Rees 1999, hereinafter MR99; SP99).  Although it contains a comparable
amount of internal energy to that of the FS, the RS bears a
temperature which is significantly lower (typically by a factor of
$\Gamma$).  As it crosses the shell, the RS produces a single short
burst of emission which is predicted to peak in the optical band with
reasonable assumptions for the burst physical parameters (MR99). The
time of peak emission, $t_{\rm peak}$, strongly depends on the initial
bulk Lorentz factor of the burst. For high $\Gamma_0$ the peak time is
comparable to the duration of the burst while for low $\Gamma_0$ it
typically occurs at later times.  More specifically, the peak time is
defined as
\begin{equation}
t_{\rm peak} = \max[\Delta/c,~t_{\rm dec}],
\end{equation}
where $\Delta$ is the width of the shell. Here, $t_{\rm dec}$, is
the time at which the inertia of the swept-up matter significantly
decelerates the shell ejecta
\begin{equation}
t_{\rm dec} = \left({3 E \over 32\pi \Gamma_0^8 n_0 m_p c^5}\right)^{1/3},
\end{equation}
where $E$ is the isotropic energy of the burst and $n_0$ characterises
the density of the external medium (SP99). $\Delta$ can be inferred
directly from the observed burst duration ($t_{\rm dur}$) by noting
that $\Delta=ct_{\rm dur}/(1+z)$ and assuming the shell does not
undergo significant spreading (Piran 1999). At a given time
(e.g. $r_{\rm peak}$), the broadband RS spectrum can be described by
the ordering of the three synchrotron break frequencies: the
self-absorption frequency $\nu_a$, the cooling frequency $\nu_c$ and
the characteristic synchrotron frequency $\nu_m$.  For spectra with
$\nu_a < \nu_m < \nu_c$, the RS is adiabatic while for $\nu_a < \nu_c
< \nu_m$ the RS is radiative, and so the electrons are cooled quickly. 
Although a radiative RS could demonstrate a peak flux of
comparable brightness to an adiabatic blast wave, the emission would
be rapidly quenched making it exceedingly difficult to detect.

\subsection{Relativistic {\it versus} subrelativistic}
  
The RS spectral break frequencies are easily calculated by comparing
them to those of the FS (MR99; SP99; Panaitescu \& Kumar 2000).  By
assuming equality of velocity and pressure across the contact
discontinuity separating the shocks, it is possible to define the
properties of the reverse shocked region in terms of $n_0$ and
$\Gamma_0$ (Blandford $\&$ McKee 1976; hereinafter BM).  However,
unlike the FS, the RS is not always relativistic (Sari \& Piran
1995). Shells satisfying
\begin{equation}
\xi \approx \left({E \over n_0 m_p c^2}\right)^{1/6} \times
\Delta^{-1/2} \Gamma_0^{-4/3} >> 1
\end{equation}
are thin and are likely to have a Newtonian RS, which is
typically too weak to decelerate the shell effectively.  On the
contrary, for thick shells ($\xi << 1$), the RS is
relativistic and thus successfully described by BM solution.
For an adiabatic blast wave peaking at $\nu_m$, the
thin shell spectral energy equations are given by
\begin{equation}
\nu_m = 5.8 \times 10^9 \epsilon_{e,-1}^2 \epsilon_{B,-2}^{1/2}
n_0^{1/2} \Gamma_0^2 (1+z)^{-1/4} {\rm Hz}
\end{equation}
\begin{equation}
F_{\nu,{\rm max}} = 4.2 \times 10^{-5} D_{28}^{-2} \epsilon_{B,-2}^{1/2} E_{50} 
n_0^{1/2} \Gamma_0 (1+z)^{3/8} {\rm Jy},
\end{equation}
while for the thick shell case we have
\begin{equation}
\nu_m = 1 \times 10^{8} \epsilon_{e,-1}^2 \epsilon_{B,-2}^{1/2} 
n_0^{1/2} \Gamma_0^2 (1+z)^{-1/2} {\rm Hz}
\end{equation}
\begin{equation}
F_{\nu,{\rm max}} = 6.0 D_{28}^{-2} \epsilon_{B,-2}^{1/2}
E_{50}^{5/4} n_0^{1/4} \Gamma_0^{-1} t_{\rm dur}^{-3/4} (1+z)^{1/2}  {\rm Jy},
\end{equation}
where $F_{\nu,{\rm max}}$ is the spectral peak flux, 
$\epsilon_e$ and $\epsilon_B$ are the equipartition functions for
the electrons ($e$) and for the magnetic field ($B$) respectively, and
$D$ is the luminosity distance of the burst.  Here we adopt the
convention $Q = 10^x\,Q_x$ for expressing the physical parameters.  It
is important to note that in the thin shell regime $F_{\nu,{\rm max}}$ scales
directly with $\Gamma_0$ while for the thick shell regime the relation
is inverted.

\subsection{Light curves of the reverse shock emission}
Unlike the synchrotron spectrum, the afterglow light curve at a fixed
frequency strongly depends on the hydrodynamics of the relativistic
shell, which determines the temporal evolution of the break
frequencies.  In a mildly relativistic RS ($\xi > 1$), the temperature
of the shocked material is non-relativistic and so the late-time
evolution of the ejecta no longer follows the BM solution.
Furthermore, the Sedov-Taylor solution is not applicable due to the
the relativistic bulk Lorentz factor of the fluid (Kobayashi \& Sari
2000; hereinafter KS00).  The hydrodynamic evolution of a mildly
relativistic RS therefore lies in a regime for which there are no
analytic solutions available.  In order to constrain the evolution of
$\Gamma$ in this regime it is common to assume (MR99; KS00)
\begin{equation}
\Gamma \propto R^{-g}~~{\rm with}~~(3/2 \le g \le 7/2)
\end{equation} 
The limits on $g$ reflect two evolutionary pathways of the pressure.
Adiabatic expansion ($p\propto \rho^{4/3}$) is assumed for $g=3/2$
while $g=7/2$ corresponds to the case of pressure equilibrium,
i.e. when the pressure of the FS matches that of the ejecta.  Using
the relation $t\propto R/\Gamma^2 c$, one can obtain the scaling of
$\Gamma$ in terms of the observer-frame time, which is given by
\begin{equation}
\Gamma \propto t^{-g/(1+2g)}.
\end{equation}
The pressure and density then scale as
\begin{equation} 
p\propto t^{-8(3+g)/7(1+2g)} \\
\rho \propto t^{-6(3+g)/7(1+2g)}.
\end{equation}
Application of these scalings to a slow cooling RS gives the spectral
evolution as a function of $g$.  For an adiabatic blast wave, it can
be shown that $\nu_m \propto \Gamma p^{5/2} \rho^{-2}$ and $F_{\nu_m}
\propto \Gamma p^{1/2}$ (KS00). In terms of the observer-frame time,
these scalings are then
\begin{equation}
\nu_m \propto t^{-3(8+5g)/7(1+2g)} \\
F_{\nu_{m}} \propto t^{-(12+11g)/7(1+2g)}.
\end{equation}
Following the peak optical emission, with the passage of time, the
light from the RS shifts rapidly to lower frequencies.  After the
RS peak frequency $\nu_m$ crosses the observing frequency it
decays rapidly. The evolution of the radio flux for frequencies
above $\nu_m$ is typically found by assuming a distribution of injected
electrons with power law of index $p$ above a minimum Lorentz
factor $\gamma_i$.  The corresponding spectral flux at a given
frequency above $\nu_m$ is then given by $F_\nu \approx F_{\nu_m}(\nu
/ \nu_m)^{-(p-1)/2}$; with $F_\nu \propto \nu^2$ at low frequencies
($\nu < \nu_a$) and $F_\nu \propto \nu^{1/3}$ bridging the two
regimes. Together with the scalings of equation 11, the above 
expression gives the evolution of the radio flare immediately after
the $\nu_m$ crossing:
\begin{equation}
F_{\nu} \propto T^{-(7+24p+15pg)/14(1+2g)} \;\;\; \nu\:>\;\nu_m
\end{equation}
It should be noted that for a typical spectral index $p=2.5$, the flux
decay index varies in a relatively narrow range ($\approx 0.4$)
between the theoretical limits of $g=[1/2,3/2]$.  Since the flux decay
index is a monotonic function of $g$ it is not very sensitive to it,
and consequently the decay shape is more strongly affected by the
value of $\Gamma_0$ than of $g$.  As demonstrated in Section 3, the
quality of GRB radio observations is not yet sufficient to
place robust constraints on $g$, and thus the hydrodynamic evolution
of the RS is currently not well determined.

\section{Constraining the kinematics and hydrodynamical evolution of the reverse shock} 

Despite ongoing observational searches, GRB 990123 remains one of few
bursts for which a radio flare has been observed.  The ability to
detect flares depends on both the strength and peak time of the
optical flash as well as the subsequent flux decay index.  In turn,
these observables depend on the ambiguous dynamics of the reverse
shock.  By constraining the kinematic and hydrodynamic properties of
the shock, we may enable observational strategies for radio flare
detection to be improved.  Moreover, as we will demonstrate,
constraints on the physics of the RS may provide us with environmental
diagnostics of GRBs.  In this section, we present a method to
determine the properties of the RS through modelling of the early
radio afterglow observations.  We constrain both the velocity of
source expansion $\Gamma_0$ and the evolutionary index $g$ of the RS
in bursts for which multi-wavelength observations have been able to
determine the physical parameters of the FS.  We use all 6 bursts
(980519, 990123, 990510, 991208, 991216, 000418) which are found to be
adiabatic (i.e. $\nu_m < \nu_c$) for reasonable assumptions about
$\Gamma_0$.  Assuming that the burst equipartition values for the FS
are shared by the RS, we apply the formalism developed in the previous
section to determine the spectral and temporal evolution of the RS. We
estimate the strength of the RS contribution in the radio band during
the epoch of afterglow observations.  Through combining our radio
flare estimates with predictions for the FS emission (obtained by
PK02), we are able to estimate the combination of $\Gamma_0$ and $g$
which provides the best $\chi^2$ fit to the early radio data. We
consider both cases where the temperature of the shocked shell is
relativistic or not. This is particularly important since for
reasonable assumptions about the velocity of source expansion we find
in all cases that $\xi \sim 1$.  For $\chi^2_{\rm FS + RS
}/\chi^2_{\rm FS} < 1$, the fit is improved by including emission from
the RS.  We will show that although current RS theory predicts a
detectable radio flare component for a wide range of $\Gamma_0$ and
$g$ parameter values, it appears to be observationally supported for
only some bursts.  Below we discuss the radio emission and RS modelling
of the individual GRBs.  A summary of results are listed in Table 1.

\begin{figure}
{\vfil
\psfig{figure=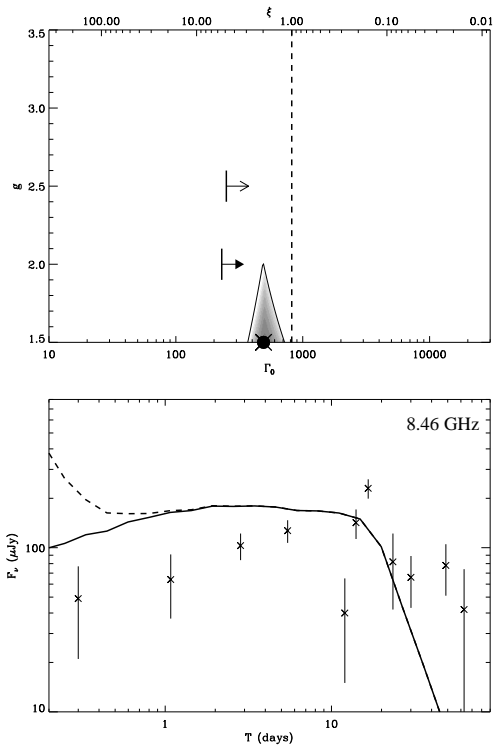,angle=0,width=0.48\textwidth}
\caption{GRB 980519: constraints on $\Gamma_0$ and $g$ assuming a thin
shell. The best fit value is marked by a solid dot and the approximate
1 $\sigma$ uncertainty contour is shaded.  A RS contribution can be
minimised for values of $\Gamma_0\sim 490$ and
$g\sim 1.5$.  Thick shell equations provide a lower limit of
$\Gamma_0 > 230$ (filled arrow) which is consistent with $\Gamma_0 >
250$ derived by PK02.  This RS is found to be mildly
relativistic. Afterglow observations are fit better with a FS model
(lower panel, solid line) than with a FS + RS model (lower panel, dashed line).}
\vfil}
\label{980519}
\end{figure}

\subsection{GRB 980519}
The broadband modelling of the FS emission by PK02 was found to
overestimate the radio observations at early times.  The addition of a
RS contribution to the FS emission serves only to worsen the fit.
Assuming the thin regime solution, we find that a minimum flux
contribution is expected from the RS for $\Gamma_0\sim 490$ and $g\sim
1.5$. Thick shell solutions for this burst do not converge since the
predicted radio flare decreases indefinitely with increasing
$\Gamma_0$. Therefore, we can only determine a lower limit of
$\Gamma_0 > 230$ for this regime.  These constraints imply a mildly
relativistic RS with $\xi \sim 1$. In this case, neither the thick nor
thin regime are ruled out by the observations. For these values of
$\Gamma_0$ and $g$, the radio observations are fit with a FS + RS
emission model giving a similar $\chi^2$ to that of the FS alone.  We
find a reduced $\chi^2=12$ and a ratio of of the quality of the fits
gives $\chi_{\rm FS+RS}^2 / \chi_{\rm FS}^2 = 1.4$ for the thin shell
case (see Figure 1). It should be noted that the $\chi^2$ ratio can in
principle decrease for values of $\Gamma_0 > 1000$ and $g=1.5$,
however, with a peak time of only a few seconds, this solution is
clearly inconsistent with the constraints given by equation
1. Therefore, by assuming the results for $\chi_{\rm
FS+RS}^2/\chi_{\rm FS}^2 = 1.4$ as our best fit, we find this solution
to be in agreement with the previous estimate by PK02 of $\Gamma_0 >
250$.

Here we comment on the structure of the $\chi^2$ region for this burst
as a representative of the larger sample included within this
study. First, it should be noted that the unusual shape of the
$\chi^2$ region is the result of an intrinsic coupling between the
fitted parameters, $\Gamma_0$ and $g$.  Although a more proper
treatment of the uncertainty region would include a full Monte Carlo
analysis to determine the error region, this level of complexity is
not warranted by the quality of available data.  We therefore apply
standard $\chi^2$ fitting techniques to build two dimensional
uncertainty regions and emphasize that these represent only
approximate confidence regions.  Current theory dictates that the
uncertainty region for $\Gamma_0$ has a lower boundary of $\Gamma_0
\sim 30$ due to the lack of photon-photon attenuation (M\'esz\'aros,
Laguna \& Rees 1993) and an upper boundary of $\Gamma_0\sim 10^3$
based on the pulse width evolution of the prompt emission
(Ramirez-Ruiz \& Fenimore 2000).  These theoretical bounds should be
adopted as the overall uncertainty region for each of the bursts
within this sample.  Secondly, it should be noted that the sharp
inflection points traced by the confidence region are an artifact of
the discontinuity imposed by the transition from a non-relativitsic to
a relativistic regime.  The tendency for the best-fit value of
$g$ to favour the parameter space extrema demonstrates that $\Gamma_0$
is the dominant parameter in this model and thus the Lorentz factor
drives the spectral evolution of the RS emission within this parameter
space (as discussed in Section 2.2).

\begin{figure}
\psfig{figure=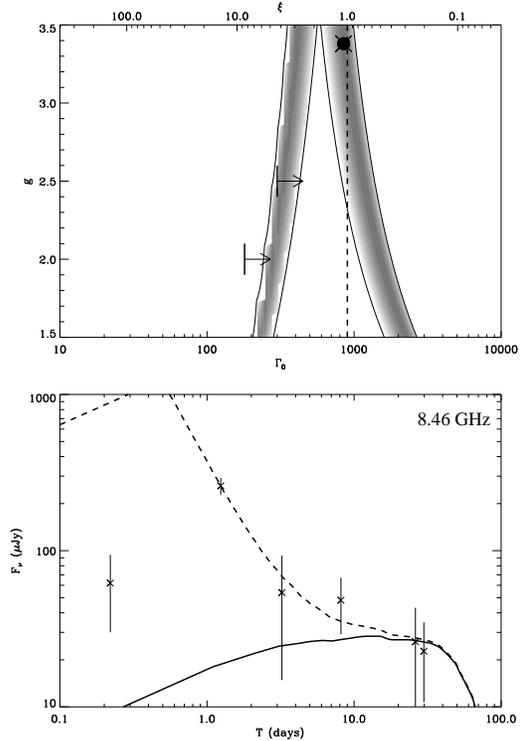,angle=0,width=0.48\textwidth}
\caption{GRB 990123: Constraints on $\Gamma_0$ and $g$ assuming a thin
shell regime. Best fit values are marked by a solid dot and approximate 1
$\sigma$ uncertainty contours are shaded. Radio flare parameters which
produce the best fit to the data $\Gamma_0\sim 850$ and $g\sim 3.4$ for
the thin shell case.  Best fit FS + RS light curve models are plotted
(dashed line) within the lower panel in order to compare with FS model
alone (solid line).  The FS+RS model provides a significantly better fit
to the radio emission.  (Note the discrepancy associated
the first point is due to the effects of self-absorption.)  The
$\Gamma_0$ constraints are consistent with the estimate of $\Gamma_0 >
300$ (upper arrow) derived by PK02 and the lower limit of $\Gamma_0 >
180$ (lower arrow) by LS01.  Note the mildly relativistic nature of
the thin shell regimes, with $\xi \sim 1$.}
\label{990123a}
\end{figure}

\begin{figure}
{\vfil
\psfig{figure=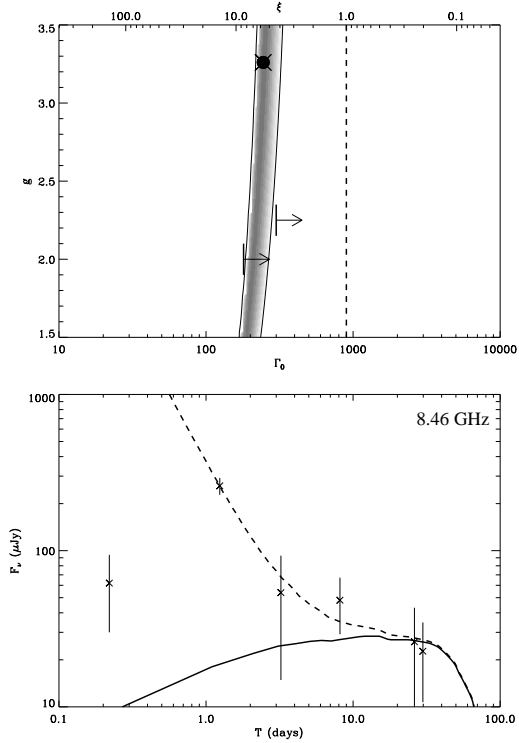,angle=0,width=0.48\textwidth}
\caption{GRB 990123: Constraints on $\Gamma_0$ and $g$ assuming a
thick shell regime. Best fit values are marked by a solid dot and
approximate 1 $\sigma$ uncertainty contours are shaded. Radio flare
parameters which produce the best fit to the data $\Gamma_0\sim 240$
and $g\sim 3.3$ for the thick shell case.  Best fit FS + RS light
curve models are plotted (dashed line) within the lower panel in order
to compare with FS model alone (solid line).  It should be noted that
the thin shell equations produce better fits to both radio and optical
emission (see Figure 2).  The $\Gamma_0$ constraints are consistent
with the estimate of $\Gamma_0 > 300$ (upper arrow) derived by PK02
and the lower limit of $\Gamma_0 > 180$ (lower arrow) by LS01.  Note
the mildly relativistic nature of the thick shell regimes, with $\xi
\sim 6$.}  \vfil}
\label{990123b}
\end{figure}

\subsection{GRB 990123}
Early radio observations of GRB 990123 revealed a flare rising to
maximum after $\sim 1$ day and then rapidly fading away (Kulkarni et
al. 1999). The strong dimming of the radio emission after $\sim 2$
days excluded a FS origin. The simplest interpretation was that it
arises from the RS (Kulkarni et al. 1999; Panaitescu \& Kumar 2000).
As expected, the FS modelling by PK02 significantly underestimates the
radio emission at the time of the flare.  By including a RS
contribution, a FS + RS emission model improves the fit by a factor of
$\sim 100$.  Best fit values are obtained for
$\Gamma_0\sim 850$ ($\Gamma_0\sim 240$) and
$g\sim 3.4$ ($g\sim 3.3$) for the thin (thick)
shell case (Figures 2 and 3).  
Both regimes predict a mildly relativistic RS
with $\xi$ values of $\sim$ 1 and $\sim$ 6 for the thin and thick
shell solutions, respectively.  Despite the differences between the
regimes, both solutions show a minimum at $\chi^2_{\rm FS
+RS}=0.3$ which gives a ratio  $\chi^2_{\rm FS+RS}/\chi^2_{\rm FS} 
\sim 0.01$, thus demonstrating a remarkable improvement to the FS fit.  
Note that the inclusion of self-absorption effects, which are not
included in the FS model of PK01 or the FS+RS model presented here,
may enable a better description of the radio observations before
one day (the reader is referred to SP99 for this matter). 
It should be commented that within the $1
\sigma$ uncertainty region of the thin regime solution, there is a
secondary minimum which occurs for lower Lorentz factors, near
$\Gamma_0\sim 400$ and $g\sim 3.5$. This solution corresponds to a peak time of
200 seconds which is clearly inconsistent with the observed optical
light-curve which peaks at $\sim 50$ seconds (Akerlof et al. 1999).
Based on the observed behaviour of the optical flash, which is
successfully described by a mildly relativistic shell (see Kobayashi
2000), we thus conclude that the thin shell solution is the regime which
better describes the RS behaviour.  We find the values of $\Gamma_0$
consistent with previous estimates for this burst: $\Gamma_0\approx
1200$ (Wang, Dai $\&$ Lu, 2000), $\Gamma_0=1400\pm 700$ (Panaitescu
$\&$ Kumar, 2001a), $\Gamma_0\approx 900\pm 100$ (SR02), $\Gamma_0 >
180$ (Lithwick $\&$ Sari 2001, hereinafter LS01), and $\Gamma_0 > 300$
(PK02).  Estimates for the dimensionless thickness parameter, $\xi$,
are similarly in agreement and include: $\xi\sim 0.7$ (KS00) and
$\xi=0.99$ (SR02).  Furthermore, the hydrodynamical evolution has
previously been characterised by $g=2.2$ (KS00) which is consistent
within our approximate 1 $\sigma$ uncertainty region.

\begin{figure}
{\vfil
\psfig{figure=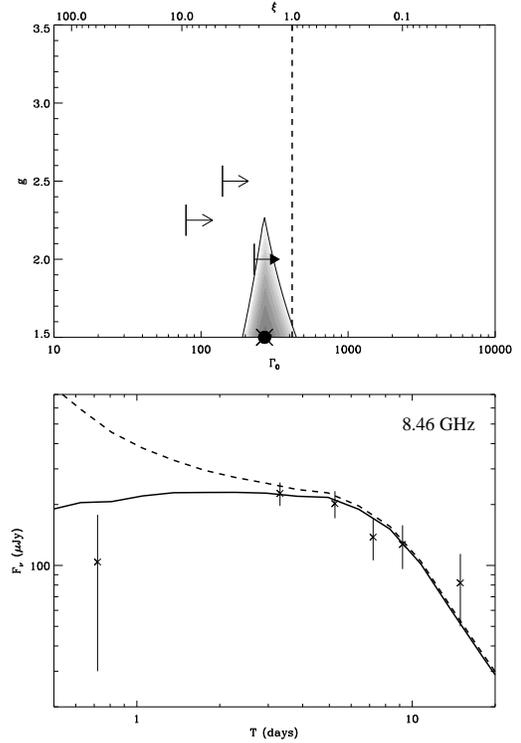,angle=0,width=0.48\textwidth}
\caption{GRB 990510: Constraints on $\Gamma_0$ and $g$ assuming a thin
shell.  The best fit value is marked by a solid dot and the approximate 1
$\sigma$ uncertainty contour is shaded.  The radio flare is minimised
for values of $\Gamma_0\sim 270$ and $g\sim 1.5$ with
$\xi \sim 2$.  Thick shell equations provide a lower limit of $\Gamma_0
> 230$ (filled arrow) which is consistent with $\Gamma_0 > 140$
derived by PK02 (upper unfilled arrow) and $\Gamma_0 > 79$ by LS01 (lower unfilled arrow).  Afterglow observations are better
fit with a FS model (lower panel, solid line) than with a FS + RS model
(lower panel, dashed line).}  
\vfil}
\label{990510}
\end{figure}

\subsection{GRB 990510}
As in the case of GRB 980519, the broadband modelling of the FS
emission of this burst overestimates the radio observations at early
times (PK02).  We find that a minimum flux contribution from the RS
can be attained from the thin shell equations by setting
$\Gamma_0\sim 270$ and $g\sim 1.5$.  The thick shell
solution provides only a constraint of $\Gamma_0 > 230$.  This RS is
found to be mildly relativistic with $\xi \sim 1$.  A FS + RS model
gives a best fit value with $\chi^2=8.5$ and  $\chi^2_{\rm FS +
RS}/\chi^2_{\rm FS}\sim 8.0$ indicating that the standard FS fit is
still preferred (see Figure 4).  A better $\chi^2$ ratio ($\sim 1$) can be
obtained for values of $\Gamma_0 > 2000$ and $g=1.5$.  We disregard
this solution, however, as it corresponds to $t_{\rm peak} < 1$ s.  
Note that these results are
in agreement with the estimate of $\Gamma_0 > 140$ derived by PK02
and the lower limit of $\Gamma_0 > 79$ by LS01.

\begin{figure}
{\vfil
\psfig{figure=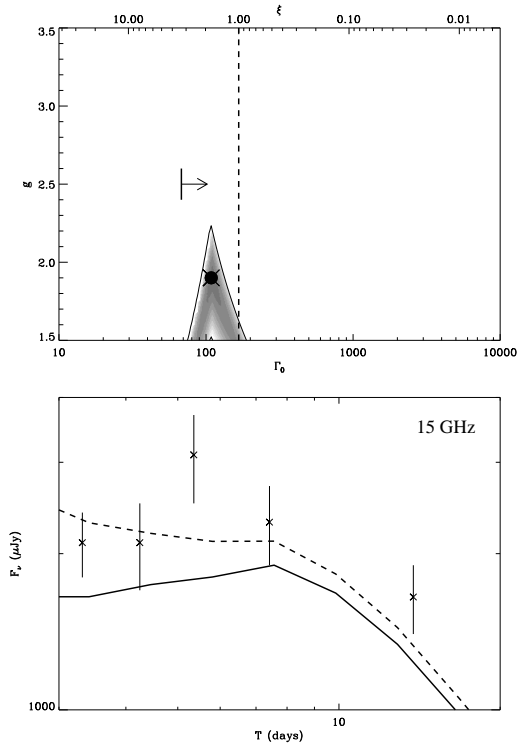,angle=0,width=0.48\textwidth}
\caption{GRB 991208: Constraints on $\Gamma_0$ and $g$ assuming a thin
shell.  The best fit value is marked by a solid dot and the approximate
1 $\sigma$ uncertainty contour is shaded.  Best parameter fits are
$\Gamma_0\sim 110$  and
$g\sim 1.9$  for  $\chi^2\sim 1.9$.  Both the thick and thin
shell regimes favour mildly relativistic solutions with $\xi \sim 1$.  The
FS + RS model (lower panel, dashed line) improves the quality of the light curve fit by
a factor of $\sim 2.5$ as compared with the FS model (lower panel, solid
line). Results are consistent with the PK02 constraint of
$\Gamma_0>68$ (arrow)}
\vfil}
\label{991208_15a}
\end{figure}

\begin{figure}
{\vfil
\psfig{figure=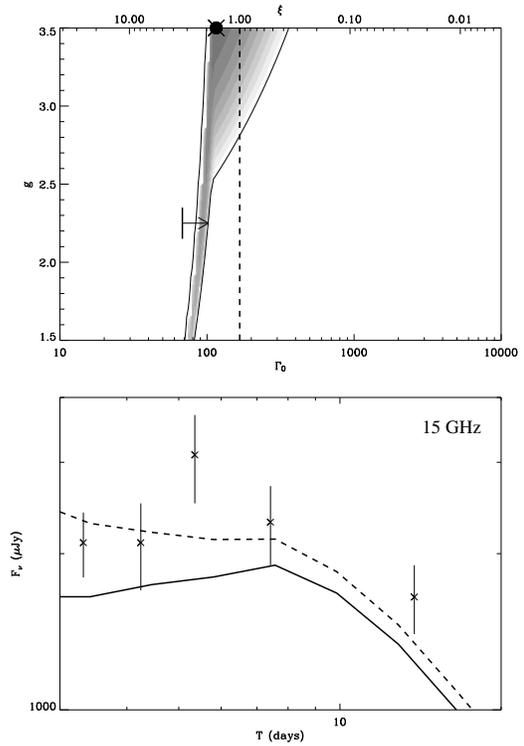,angle=0,width=0.48\textwidth}
\caption{GRB 991208: Constraints on $\Gamma_0$ and $g$ assuming a
thick shell.  The best fit value is marked by a solid dot and the approximate
1 $\sigma$ uncertainty contour is shaded.  Best parameter fits are $\Gamma_0\sim 130$ and
$g\sim 3.5$ for a $\chi^2\sim 1.5$.  Both the thick and thin shell regimes
favour mildly relativistic solutions with $\xi \sim 1$.  The FS + RS
model (lower panel, dashed line) improves the quality of the light
curve fit by a factor of $\sim 2$ as compared with the FS model
(lower panel, solid line). Results are consistent with the PK02
constraint of $\Gamma_0>68$ (arrow).}  \vfil}
\label{991208_15a}
\end{figure}

\subsection{GRB 991208}
Broadband fits to the FS of GRB 991208 included two sets of radio
lights curves observed at 8.5 GHz and 15 GHz.  We calculate radio
flare predictions for both observing frequencies.  At 15 GHz, the FS
prediction underestimates the data (PK02).  By adding a RS
contribution, we find the ratio $\chi^2_{\rm FS+RS}/\chi^2_{\rm FS}$
can be improved by a factor of $\sim 2$ for both thin and thick
solutions.  Best FS + RS fits are produced using $\Gamma_0\sim 130$
and $g=3.5$ for the thick shell equations which give $\chi^2_{\rm
FS+RS}\sim 1.5$ and a ratio of $\chi^2_{\rm FS+RS}/\chi^2_{\rm FS}
\sim 0.4$.  The thin shell solution is similar in quality, with best
values of $\Gamma_0\sim 110$ $g\sim 1.9$, giving
$\chi^2_{\rm FS+RS}\sim 1.9$ and a ratio of $\chi^2_{\rm
FS+RS}/\chi^2_{\rm FS}\sim 0.5$.  Both regimes predict a mildly
relativistic RS with $\xi \sim 1$.  At 8.5 GHz, a thin RS is unable to
provide solutions which improve the FS fit.  The best result gives
$\chi^2_{\rm FS +RS }/\chi^2_{\rm FS}\sim 1.4$ for values of
$\Gamma_0\sim 110$ and $g\sim 1.5$. Thick shell equations
fail to converge and therefore predict only a lower limit of $\Gamma_0
> 130$. Combining the two solution sets should generally enable
further constraints.  For this case, however, combining the sets
merely reproduces the 15 GHz solution since this set of results is
statistically dominant.  Therefore, we quote the 15 GHz results as our
best fits for this RS solution (Figures 5 and 6).  Notice that the results
are consistent with the PK02 limit of $\Gamma_0 > 68$.
\begin{figure}
{\vfil
\psfig{figure=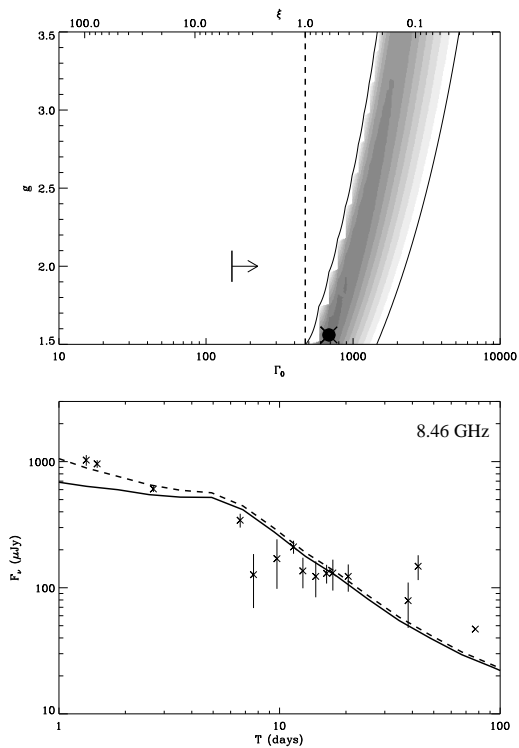,angle=0,width=0.48\textwidth}
\caption{GRB 991216: Constraints on $\Gamma_0$ and $g$ assuming a thin
shell.  The best fit value is marked by a solid dot and the approximate
1 $\sigma$ uncertainty contour is shaded.  RS parameters which produce the best fit to the data are
$\Gamma_0\sim 680$ and $g\sim 1.55$ for $\chi^2\sim 4.2$.  The
lower panel compares the best fit FS+RS light curve model (dashed line)
and the FS model alone (solid line). This value of $\Gamma_0$ is
consistent with the PK02 constraint of $\Gamma_0 > 150$ (arrow).  Note
that the thin solution favours a mildly relativistic RS.  } \vfil}
\label{991216a}
\end{figure}

\begin{figure}
{\vfil
\psfig{figure=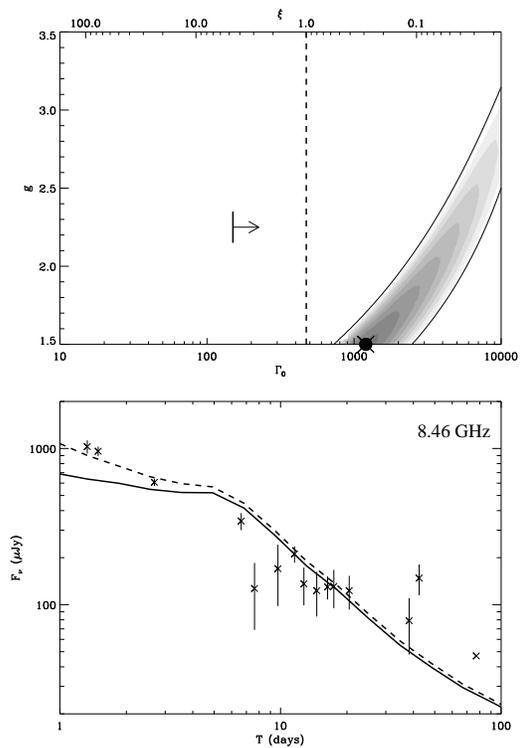,angle=0,width=0.48\textwidth}
\caption{ GRB 991216: Constraints on $\Gamma_0$ and $g$ assuming a
thick shell.  The best fit value is marked by a solid dot and the approximate
1 $\sigma$ uncertainty contour is shaded. RS parameters which produce the best fit to the data are
$\Gamma_0\sim 1200$ and $g\sim 1.5$ for a thick shell.  The lower
panel compares the best fit FS+RS light curve model (dashed line) and
the FS model alone (solid line).  This value of $\Gamma_0$ is
consistent with the PK02 constraint of $\Gamma_0 > 150$ (arrow).  In
comparison with the thin shell constraint, the thick solution tends
towards the relativistic limit with $\xi<1$.}  \vfil}
\label{991216b}
\end{figure}
\subsection{GRB 991216}
Early radio observations of GRB 991216 suggest the emission was
already in decline after $t\sim 1$ day (Frail et al. 2000).  As a
result, efforts to fit the radio light-curve with a standard FS
emission model proved difficult. Broadband fits by Panaitescu \& Kumar
(2000) underestimate the early radio observations and claim that
interstellar scintillation is essential in explaining the departures
between observations and model fluxes.  Alternative models have also
been suggested, including a dual fireball, and a RS flare (Frail et
al. 2000, Panaitescu \& Kumar 2000).  We find that the fit can be
improved by a factor of $\sim 2$ by including emission from the RS.
Both thick and thin shell solutions give $\chi^2_{\rm FS+RS}\sim 4.2$.
The thin shell solution favours a mildly relativistic RS with best fit
values of $\Gamma_0\sim 680$ and $g\sim 1.55$ (Figure 7) while the thick
shell solution gives a highly relativistic solution with $\Gamma_0\sim
1200$ and $g\sim 1.5$ (Figure 8).  We note that this solution corresponds
to a peak time less than $\sim 2$ sec which is considerably smaller
than the estimated light width of the shell $\sim 15$ s. Both thick
and thin regimes produce RS solutions which are consistent with the
PK02 constraint of $\Gamma_0 > 150$.
\begin{figure}
{\vfil
\psfig{figure=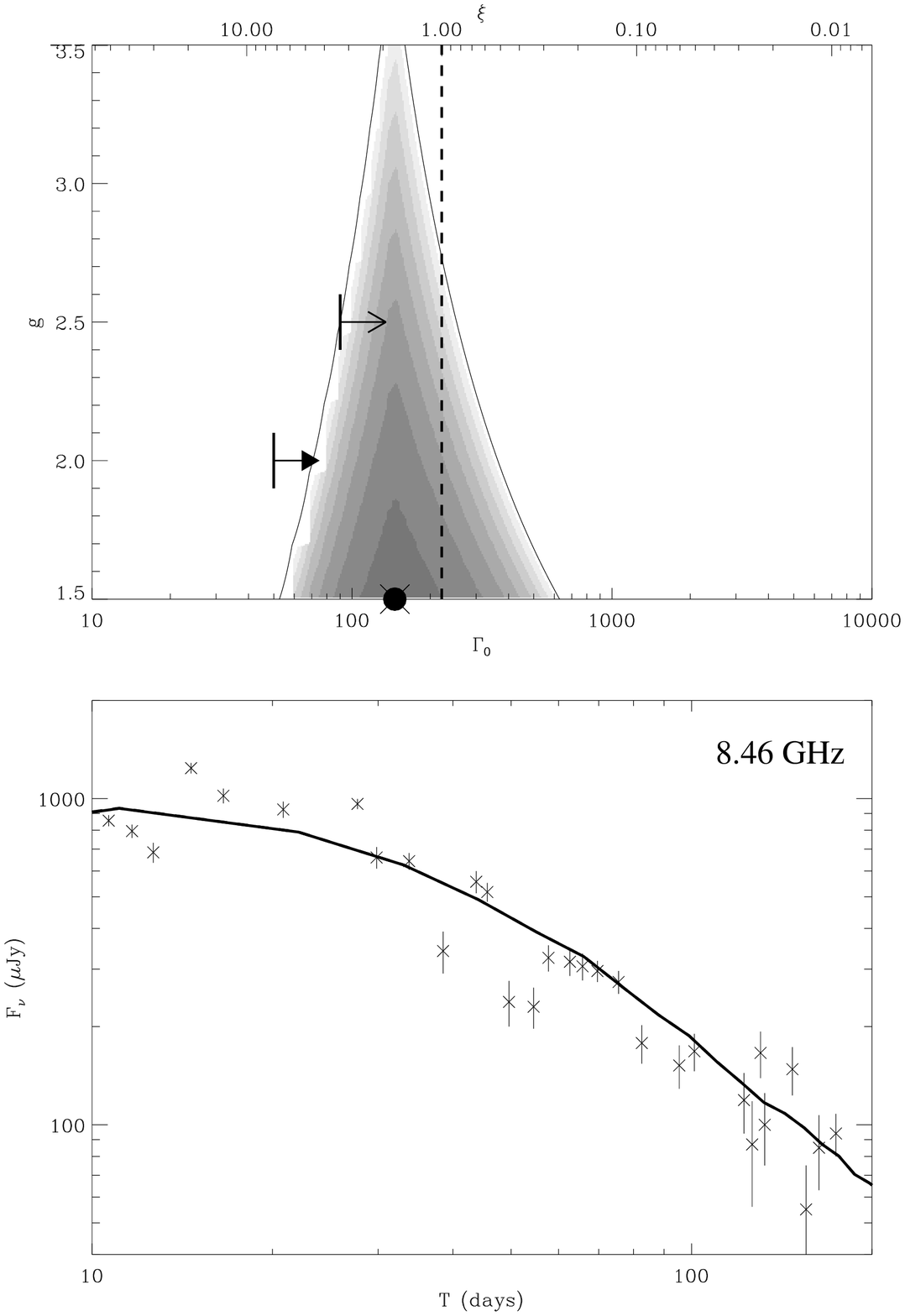,angle=0,width=0.48\textwidth}
\caption{GRB 000418: Constraints on $\Gamma_0$ and $g$ assuming a thin
RS.  The best fit value is marked by a solid dot and the approximate 1 $\sigma$
uncertainty contour is shaded.  A RS contribution can be minimised for
values of $\Gamma_0\sim 150$ and $g\sim 1.5$.  Thick shell equations
provide a lower limit of $\Gamma_0 > 50$ (filled arrow) which is less
constraining than $\Gamma_0 > 90$ as derived from jet modelling
(PK02).  This shock is mildly relativistic with $\xi \sim 1.7$.  The
lower panel compares the resulting best fit FS+RS model light curve
(dashed line) with the FS model (solid line).}  \vfil}
\label{000418}
\end{figure}

\subsection{GRB 000418}
Broadband afterglow fits to the FS emission are in agreement with the
radio observations (Berger et al. 2001; PK02).  It is clear that the
addition of a RS contribution will not improve the fit to the radio
light-curve.  A minimum flux contribution is expected from the RS for
$\Gamma_0\sim 150$ and $g\sim 1.5$ assuming a thin shell
(Figure 9).  This implies a mildly relativistic RS with $\xi \sim 2$.
The radio flux in the thick shell solution decreases with increasing
bulk Lorentz factor and thus we can only determine a lower limit of
$\Gamma_0>50$ for this regime. It is emphasised that the $\chi^2$
ratio never falls below 1 (with the best fit $\chi^2_{\rm FS+RS}\sim 9.7$)
for all values of $\Gamma_0$ and $g$. It also should be noted that the
relatively large uncertainty region for this burst is due to the
predicted faintness of the radio flare at the epoch of observations
($\sim 10$ days).

\begin{table*}
 \centering
\caption{Constraints on $\Gamma_0$ and $g$ obtained by fitting a RS + FS
model to early radio observations.  The reduced $\chi^2$ for the best FS+RS
model is given as well as the ratio $\chi^2_{\rm FS + RS }/\chi^2_{\rm FS}$.  For $\chi^2_{\rm FS + RS }/\chi^2_{\rm FS} < 1$, the fit is
improved by including emission from the RS.}

\begin{tabular}{|l|rrrr|rrrr} \hline\hline
\large \sc GRB & \multicolumn{4}{c}{ \large \sc Thin regime} &
\multicolumn{4}{c}{ \large \sc Thick regime} \\ 
\normalsize 
& $\Gamma_0$ & $g(g_{low}-g_{high})$ & $\chi^2_{\rm FS+RS}$ & $\chi^2_{\rm FS+RS}/\chi^2_{\rm FS}$
& $\Gamma_0$ & $g(g_{low}-g_{high})$ & $\chi^2_{\rm FS+RS}$ & $\chi^2_{\rm FS+RS}/\chi^2_{\rm FS}$ \\ \hline

\bigskip
 
 980519\footnote{redshift unknown. z=1 was assumed.} & $490$ & 1.5 & 12 & 1.4 & $>$230 & 1.5 & - & - \\

\bigskip

 990123 & $850$ & 3.4 & 0.3 & 0.01 & $240$ & 3.3 & 0.3 & 0.01\\

\bigskip

 990510 & $270$ & 1.5 & 8.5 & 8.0 & $>$230 & 1.5 & - & - \\

\bigskip

 991208 & $110$ & 1.9 & 1.9 & 0.5 & $130$ & 3.5 & 1.5 & 0.4\\ 

\bigskip

 991216 & $680$  & 1.5  & 4.2 & 0.6 & $1200$ & 1.5 & 4.2 & 0.6\\

 000418 & $150$ & 1.5 & 9.7 & 1.0 & $>$50 & 1.5 & - & - \\ \hline\hline

\end{tabular}
\end{table*}

\subsection{Caveats}

A number of caveats apply to our analysis. First and foremost, the
hydrodynamical and kinematical constraints presented here have been
calculated under the assumption that the radio flare arises from the
reverse shock component of the blast wave (MR99 and SP99). Within the
grounds of the FS model, the radio flare (and  in
particular those observed in 990123, 991216 and 991208) can be
accounted for only if the characteristic frequencies can be evolved
much faster than that given by standard dynamics (as argued also by
Kulkarni et al. 1999 for 990123). In contrast, the
RS model provides a natural and consistent explanation for
the radio flare. 

In the
framework of ``standard'' afterglow models - by which we mean models that
assume a power-law-shocked particle spectrum and a constant fraction
of energy in electrons and magnetic fields relative to the thermal
energy of the shocked particles - we have estimated the RS spectra by
assuming that the equipartition values for the FS component are
unchanged across the contact discontinuity. Although the validity of
this assumption is questionable, it is a convenient method of
estimating these illusive parameters until more is known about the
microphysics of GRB shocks.  We test the strength of this assumption by
altering $\epsilon_B$ (the parameter for which there is larger
uncertainty) by two to three orders of magnitude. The effect of strongly
increasing $\epsilon_B$ is only a slight shift in the $\Gamma_0$ - $g$
confidence regions. Such shifts, however, are always found to lie
within the uncertainty regions given for each burst, thus
supporting the validity of our results. 

The premise of our work has been that the physical parameters derived
by PK02 provide a robust description of the FS emission. Most
importantly, we assume that the PK02 parameters are not significantly
biased by neglecting the effects of a RS component.  This assumption
is supported by the fact that the PK02 results are weighted towards
the optical and X-ray data where the sampling frequency was large, the
observational errors were small, and the RS component was negligible on
timescales $\sim 1$ day.  With sparse observations and large
error bars, the radio data clearly does not constrain the
multi-frequency fit at the same level as the optical.  The RS
predictions presented here, for all bursts in our sample, proved to be
generally consistent with the observed light-curves, and it is
critical to note that none of the RS predictions significantly
over-estimated the observed radio emission, even at early times. Our
RS predictions therefore provide consistency checks for PK02
physical parameters, which we find to be in agreement.

\section{The elusive $\Gamma_0$}   
As discussed in SR02, the bulk Lorentz factor plays a critical role in
the majority of GRB emission models (see Piran 1999 for a review).
The Lorentz factor, however, has also proven to be one of the most
difficult physical parameters to constrain from gamma-ray burst
observations.  Based on theoretical predictions and sparse
observations, present estimates stretch over two orders of magnitude:
$\Gamma_0\approx 10-10^{3}$ (e.g. M\'esz\'aros, Laguna \& Rees 1993;
Wang, Dai $\&$ Lu 2000; Ramirez-Ruiz \& Fenimore 2000; PK02; LS01; SR02).  
Here we present
constraints on $\Gamma_0$ for 6 of the best-sampled bursts observed to
date.  Although the $\Gamma_0$ constraints are still quite broad in
some cases (e.g. 000418), information may still be gained from the
collective $\Gamma_0$ distribution for our data set.  As a method of
investigating this, we treat each burst's constraint as a probability
distribution and produce a total histogram by summing over the 6
normalised distributions.  Specifically, for each burst, we construct
a one dimensional probability histogram for $\Gamma_0$, given by
$p_i(\Gamma)d\Gamma$.  Here, $p_i(\Gamma)d\Gamma$ represents the
frequency of each $d\Gamma$ value within the 1-$\sigma$ $\Gamma_0-g$
confidence region.  The $p_i(\Gamma)d\Gamma$ probability histograms
are each normalised to unity before summing them to create a total
probability histogram, constructed as $p(\Gamma)d\Gamma=\sum_i
p_i(\Gamma)d\Gamma$.  Figure 10 plots the total probability histogram
for the data set as a shaded and smoothed curve.  A Gaussian fit to
the total histogram is over plotted with a mean value of
$\Gamma_0=150$.  It should be noted that the distribution
is weighted heavily to the $\Gamma_0$ constraints derived for GRB
991208, since this burst was the most tightly constrained within the
data set (due to additional observations at 15 GHz).  To remove this
burst from the total probability histogram leaves 5 bursts with radio
flare constraints derived from 8.46 GHz data only.  The total
probability histogram associated with this smaller set of bursts has a
larger mean bulk Lorentz factor which may represent the collective
constraints more accurately, with $\Gamma_0=270$.  We
emphasise that this mean value is consistent with lower limits on
$\Gamma_0$ as derived by LS01 for a set of 12 bursts and also with
estimates for $\Gamma_0$ by PK02 for 10 bursts.  As additional
$\Gamma_0$ constraints are gathered from future bursts, the
distribution of Lorentz factors should offer vital clues concerning
the diversity of GRB source expansion velocities.

\begin{figure}
{\vfil
\psfig{figure=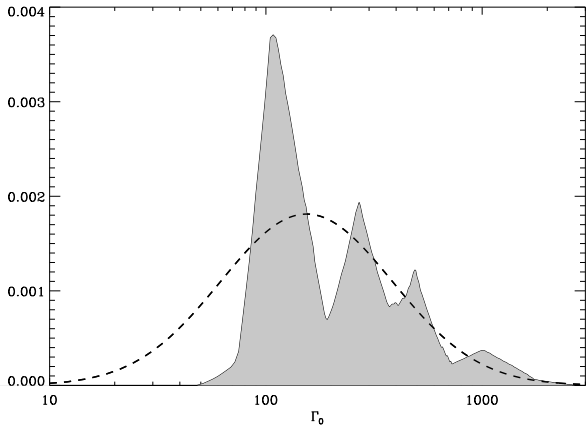,angle=0,width=0.48\textwidth}
\caption{Distribution of GRB Lorentz factors. This distribution is
essentially a smooth histogram of the data, but one that takes into
account the uncertainties in the measurements. The dashed curve is the
distribution under the curve (shaded) but smoothed with a Gaussian
with a mean value of $\Gamma_0=150$. }  
\vfil}
\label{histogram}
\end{figure}

\section{GRB environments and progenitors}

Models for the GRB afterglows indicate that the emission comes from a
region $\sim 10^{16} -10^{18}$ cm from the source of the
explosion. The nature of the material depends on the GRB progenitors,
which are at present not known.  The question of a wind versus a
constant density medium is a crucial one, since massive stars, one of
the leading candidates for GRB progenitors (e.g. Woosley 1993; Paczynski
1998), should be surrounded by a $\rho \propto r^{-2}$ wind
(unless a supernova explosion occurs before the burst; Vietri \&
Stella 1998; Konigl \& Granot 2002; Guetta \& Granot 2002). In
contrast, GRBs resulting from compact star mergers, the other leading
candidate, are expected to be surrounded by the interstellar medium
(e.g. Lattimer \& Schramm 1976; Paczy\'{n}ski 1986; Narayan, Paczy\'{n}ski
\& Piran 1992; Perna \& Belczynski 2002; Rosswog \& Ramirez-Ruiz
2002). 

Considerable discussion has recently been given to whether the
expected $r^{-2}$ density structure for a stellar wind is compatible
with analysis of the afterglow light curves (e.g. CL00; Lazzati \&
Perna 2002; PK02; Price et al. 2002). PK02 found that in half of the cases they modelled,
a homogeneous ambient medium accommodates the afterglow emission
better than the wind-like medium. Although wind-like external medium
solutions were reported as being a poor match to the data in the other
cases, PK02 include them as possible alternatives for 3 of the 6
bursts in our sample: 991208, 991216, and 000418.  To adopt the
$r^{-2}$ solutions for these 3 bursts would place their RSs in the
radiative (fast cooling) regime.  At best, a wind medium radiative RS
can attain an optical peak flux which is comparable in magnitude to
the homogeneous medium adiabatic case.  For typical parameters,
however, the optical flux is 6 times \it fainter \rm in an wind
environment than the constant density medium case
(CL00). Specifically, the peak flux is given by
\begin{eqnarray}
F_{\nu, max} & = & 46 (1+X) {\left(\frac{1+z}{2}\right)}^{1/2}
{\left(\frac{2-\sqrt{2}}{1+z-\sqrt{1+z}}\right)}^2 \nonumber \\ & &
\times {\left(\frac{\epsilon_B}{0.1}\right)}^{-1/4}
\frac{E_{52}^{5/4}}{A_{\*}^{1/2} \Gamma_{0,3} \Delta_{10}^{3/4}} {\rm
mJy}
\end{eqnarray}
where $X$ is the fractional abundance of hydrogen ($\sim 0$ for
Wolf-Rayet winds) and $A_{\*}$ is a constant which describes the
density profile of the wind (CL00).  This peak corresponds to an
optical flash of magnitude $m_{\rm V}\approx 12$ as compared with an
$r^{-2}$ estimated peak of $m_{\rm V}\approx 9$.  Furthermore, the RS
emission dies off more rapidly when surrounded by a wind external
medium.  The ability to detect a RS contribution within early radio
observations is then generally reduced for an $r^{-2}$ burst.
Therefore, after modelling the early afterglows of 991216 and 991208,
which are better described with the addition of a RS contribution, we
argue that the homogeneous medium solution is preferred.
As a consistency check, 
one can instead assume the lower limits in the value of $\Gamma_0$ given 
by PK02 and fit for the density of the external medium.
By doing so, we find density values that are consistent with
those from the FS model, which implies that there is 
no direct evidence for large
scale clumps within the density profile.

\begin{table*}
 \centering
\caption{Wind-bubble parameter constraints for 6 bursts. Estimates
derived by requiring the expanding observed blast wave to propagate
in the low-density, (uniform) shocked wind region. Best fit parameters
of both $M_{\rm ej}$ and deceleration radius are used (see Figure 11),
which together with the inferred circumburst density give $v_w$,
$\dot{M}$ and  $n_{0}$ via equations (14), (15) and (16).} 

\begin{tabular}{|l|rrrr} \hline\hline
\large \sc GRB & $v_w$ & $\dot{M}$ & $n_{0}$ \\
& (1000 km/s) & $10^{-6} M_{\odot}$/yr & $\rm cm^{-3}$ \\ \hline \hline

\bigskip

980519 & 0.58 & 0.3 & 100. \\

\bigskip

990123 & 0.50 & 0.015 & 3.4 \\

\bigskip

990510 & 1.4 & 0.2 & 55. \\

\bigskip

991208 & 1.0 & 0.7 & $>10^4$ \\

\bigskip

991216 & 1.8  & 0.08 & $>10^4$ \\

\bigskip

000418 & 1.4 & 0.9 & $>10^4$ \\ \hline\hline

\end{tabular}
\end{table*}

If the progenitors are massive stars then there is an analogy to the
explosions of core collapse supernovae, for which there is abundant
evidence that they interact with the winds from the progenitor
stars. In most supernova cases, the radial range that is observed is
only out to a few $10^{17}$ cm, such that the mass loss characteristics have
not changed significantly during the time that mass is supplied to the
wind (CL00). The density in the wind depends on the type of
progenitor. Red supergiant stars, which are thought to be the
progenitors of Type II supernovae, have slow dense winds. Wolf-Rayet
stars, which are believed to be the progenitors of Type Ib/c
supernovae and possibly of GRBs (e.g. MacFadyen \& Woosley 1999), have
faster lower density winds. Deceleration due to this wind starts in
earnest when about half the initial energy is transferred to the
shocked matter, i.e. when it has swept up $\Gamma_0^{-1}$ times its
own rest mass. The typical mass where this happens is $M_{\rm
ej}=E/(\Gamma_0^{2}c^2) \sim 5 \times 10^{-6} E_{53}\Gamma_{0,2}^{-2}
M_\odot$ (M\'esz\'aros \& Rees 1993).

\begin{figure}
{\vfil
\psfig{figure=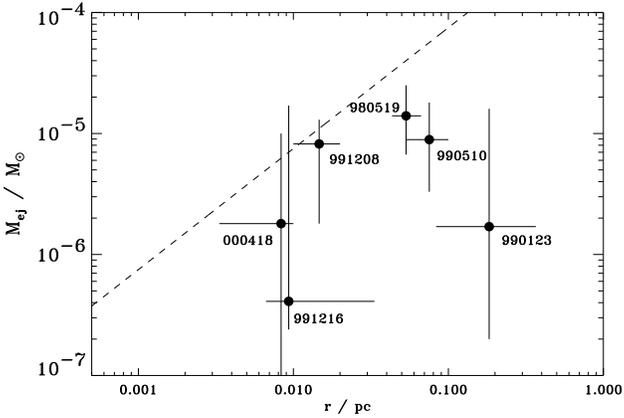,angle=0,width=0.48\textwidth}
\caption{Constraints on the swept up mass $M_{\rm ej}$ as a function
of radius. The dashed curve is the ejected stellar mass assuming
$\dot{M}=10^{-6} M_\odot {\rm yr}^{-1}$ and $v\sim 10^{3} {\rm km\;s}^{-1}$.
The errors in $M_{\rm ej}$ and radius take into account the
uncertainties in the $\Gamma_0$ estimates. }
\vfil}
\label{e_gamma}
\end{figure}
Figure 11 shows the (isotropic equivalent) swept up mass as a function
of radius derived from our $\Gamma_0$ estimates.  It should be noted,
however, that the afterglows sample a region $\le 10^{17}$ cm ($\sim
0.03$ pc) in size. Depending upon the wind history of a Wolf-Rayet
star during its last few centuries, the density structure in this
region could be quite complicated as the star enters advanced burning
stages unlike those in any Wolf-Rayet star observed so far (the
uncertainty in the evolution of massive stars leaves open the
possibility of interaction with denser material at early times; see
e.g. Ramirez-Ruiz et al. 2001). Still, there is no obvious way for the
ejected mass at $10^{17}$ cm to be much lower than about
$10^{-5}M_\odot$ (assuming $\dot{M}=10^{-6}M_\odot {\rm yr}^{-1}$ and
$v_w \sim 10^{3} {\rm km\;s}^{-1}$; see dashed line in Fig. 11). The
low swept up mass inferred is thus problematic for the collapsar
model. This has lead to the suggestion that the fireball expansion may
be taking place inside the constant density medium that is expected
downstream from the termination shock of the massive star wind (Wijers
2001; see also Scalo \& Wheeler 2001). The radius of the wind
termination shock at the inner edge of the wind bubble can be found by
balancing the wind ram pressure with the post-shock cavity pressure.
The termination shock radius, $R_t$, is thus given by
\begin{equation}
R_t=0.4 \dot{M}_{-6}^{3/10} v_{w,3}^{1/10} n_{0,3}^{-3/10} t_6^{2/5} \rm{pc}
\end{equation}
where $t_6$ is the lifetime of the star in Myr and $n_0$ is the
interstellar gas density. The density in the uniform shocked wind
region, $n_{\rm sw}$, at late times is given by
\begin{equation}
n_{\rm sw} \sim 3 \dot{M} {t \over{4\pi R^3_t} m_p} =  0.06 \dot{M}_{-6}^{4/5} n_{0,3}^{3/5} v_{w,3}^{6/5} t_6^{-4/5} \rm{cm^{-3}},
\end{equation}
which shows that even if the progenitor star is embedded in a dense
molecular cloud the observed blast wave can propagate in a
low-density, uniform medium (Wijers 2001). The mass within the
$1/r^{2}$ wind, $M_t$, is 
\begin{equation}
M_t \sim  3 \times 10^{-4} \dot{M}_6^{13/10} v_{w,3}^{-9/10} n_{0,3}^{-3/10} t_6^{2/5} M_{\odot}. 
\end{equation}
Comparison with estimates in Figure 11 show that if the wind is
especially weak (i.e. $\dot{M} \sim 10^{-6}M_\odot {\rm yr}^{-1}$) or the
surrounding pressure is high ($n_0 > 10^{3}$), $R_t$ falls within the
range of the relativistic expansion. Considering the radial
requirements relevant to our sample, we conclude that this
applies to most cases (see Table 2). Models and observations of
Galactic Wolf-Rayet stars, however, show that the swept-up shell of a
red supergiant material at the outer radius is at a distance $\ge 3$
pc from the star (Garcia-Segura et al. 1996). This radius is
sufficiently large that the interaction with the free $1/r^2$ wind is
expected over the typical period of observation of afterglows. The
calculations above demonstrate that a blast wave expanding into a wind
bubble only works for massive stars with relatively low mass winds
($\le 10^{-6}M_\odot {\rm yr}^{-1}$) and/or
embedded in dense molecular clouds $n_0 \ge 10^{3} {\rm cm}^{-3}$.\\ 

Although the interstellar and wind models are the two main types of
environments considered for afterglows, there is a different scenario
involving a massive star in which the supernova explosion occurs
before the GRB (e.g. Vietri \&
Stella 1998; Dado et al. 2002; Konigl \& Granot 2002; Guetta \& Granot
2002). The supernova would expand into the progenitor wind, creating a
complex circumburst region in the inner part of the wind. Konigl
\& Granot (2002) have recently shown, for the case of a pulsar-wind
bubble, that the shocked wind has a roughly uniform density, similar
to that found in the normal interstellar medium. 

\section{Discussion}

Accepting the inference that the radio flares arise from the reverse
shock, we place constraints on $\Gamma_0$, $g$, and hence on the
density profile of the medium in the immediate vicinity of bursts
980519, 990123, 990510, 991208, 991216, and 000418.  Unlike the
continuous forward shock, the hydrodynamic evolution of the
reverse-shocked ejecta is more sensitive. As we demonstrated, the
temperature of the reverse-shocked fluid is found to be
non-relativistic for most of these bursts. The hydrodynamics of the
{\it cold} shocked ejecta is very different from that of the {\it hot}
ejecta which is described by the BM76 solution. Surprisingly, both
cases predict rather similar light curves, with decay laws that vary
in relatively narrow ranges.  Hence, it is unlikely that the
non-detection of radio flares is linked to the relativistic regime of
their reverse shock. As argued above, the detection of such flares, or
firm upper limits, would play an important role in discriminating
between {\it cold} and {\it hot} shell evolution. \\

Moreover, the strong dependence of the peak time of this RS emission on
the bulk Lorentz factor $\Gamma_0$ provides a means to measure this
elusive parameter. Indeed, we find the fit quality of the FS + RS
model to be \it most \rm sensitive to the parameter value of $t_{\rm
peak}$.  Our $\Gamma_0$ estimates are found to typically lie between
100 and 10$^{3}$, well above the lower limits derived from the
requirement that gamma-ray bursts be optically thin to high-energy
photons (see LS01). Constrained loosely by equation
1, the peak time is predicted to have a lower limit comparable to the
light-width of the shell.  Unfortunately, the shell width can only be
approximated based on the duration of the prompt $\gamma$-ray
emission, and consequently the lower limit on $t_{\rm peak}$ involves
a significant degree of uncertainty in $\Gamma_0$.  For this sample of
bursts, the $t_{\rm peak}$ values are constrained to a large
distribution between 1 and 100 seconds. An increase in the detection
efficiency for RS flares and flashes will allow
$t_{\rm peak}$ to be better constrained. Although large localisation
errors plague attempts at early ($t < 1$ day) radio observations, we
have shown that signatures of the radio flare may still be detectable
(on top on the FS emission) at times significantly after 1 day.
Therefore, we promote the use of this constraining method in cases
where a late-time radio flare component would otherwise go unnoticed.
Certainly it is true that earlier radio observations are the key to
improved flare detection, but until current observational constraints
are lifted, we argue that it is sufficient to use observations at $t >
1$ day and still recover useful RS constraints.

Knowing $\Gamma_0$, we have then estimated the (isotropic equivalent)
swept up matter at the radius where the afterglow is produced to be
$\le 10^{-5}M_\odot$ (it should be noted that the afterglows sample a
region $\sim 0.03$ pc in size). Provisional upon the density of the
interstellar medium, this may be comparable to the wind termination
radius; but unless the the wind is especially weak or the surrounding
pressure is extraordinary high, the low densities masses inferred here
may be problematic for the collapsar model. It is evident that the
environment surrounding a massive star at the time of its death is a
very opulent one. The complex density structure we see in SN 1987A
could be a hint of what exists in some GRBs progenitors -- rich
behaviour with multiple possible transitions in the observable part of
the afterglow. If we see such transitions, they can be fairly
constraining on the properties of the progenitors. So far, the only
candidate for having shown a shock transition is the flare-up of
970508 one day after the trigger. Due to the deficiency of early data,
this interpretation is only one of many allowable.  The prompt optical
and radio emission from future bursts will give the opportunity of
investigating such transitions in detail. \\

In summary, we show the potential of modelling early radio afterglow
observations in an attempt to understand the physics of the RS. By
placing constraints on the properties of GRB shocks we hope to enable
improved observational strategies for future flash and flare
detections. We show that although current RS theory predicts a
detectable radio flare component for a wide range of $\Gamma_0$ and
$g$ parameter values, it appears to be observationally supported
for only some GRBs.  As the sample of well-observed GRB afterglows
increases, this method will become applicable to a larger set of
bursts thereby forming a more statistically significant data set.
Analysis of a larger set will offer improved diagnostics on both the
initial Lorentz factor of the collimated fireball and also the density
of the burst immediate environment.

%
\section*{Acknowledgements}
This work was motivated by conversations with and encouragement from
A. Panaitescu and P. Kumar. The authors wish to thank E. Berger, A. Blain and D. Frail for helpful comments and discussion.  
AMS was supported by the NSF GRFP. ER-R thanks 
CONACYT, SEP and the ORS for support.

\bsp

\label{lastpage}

\end{document}